\documentclass[11pt,twoside]{article}

\usepackage{times,fancyhdr}
\usepackage{cite}
\usepackage{graphicx}

\date{}

\title{Features of Traffic Congestion caused by bad Weather Conditions  or Accidents}  
\author{Boris S. Kerner\footnote{Daimler  AG, GR/ETI, HPC:  G021, 71059 Sindelfingen, Germany,
boris.kerner@daimler.com}}  

\begin{document}

\pagestyle{fancy}
\fancyhead{} % clear all header fields
\fancyhead[EC]{Boris S. Kerner}
\fancyhead[EL,OR]{\thepage}
\fancyhead[OC]{Features of Traffic Congestion caused by bad Weather Conditions  or Accidents}
\fancyfoot{} 
\renewcommand\headrulewidth{0.5pt}
\addtolength{\headheight}{2pt}

\maketitle

\begin{abstract}
 Spatiotemporal features and physics of vehicular traffic congestion 
 occurring due to  heavy freeway bottlenecks caused by bad weather conditions or accidents  are found
  based on simulations in the framework of  
 three-phase traffic theory. A model of a heavy bottleneck
 is presented. Under 
 a continuous non-limited increase in   bottleneck strength, i.e., when 
 the average flow rate within a congested pattern allowed by the  heavy bottleneck  decreases continuously up to
 zero, the evolution of the traffic phases in congested traffic,
  synchronized flow and wide moving jams,   is studied. 
  It is found that at small enough flow rate within the congested pattern, the pattern exhibits 
   a non-regular structure:
 a pinch region of synchronized flow within the pattern 
 disappears and appears randomly over time; some of the wide moving jams merge randomly into a mega-wide moving jam (mega-jam). 
 We show that the smaller the average flow rate allowed by the heavy bottleneck within the congestion, 
 the less the mean pinch region width and the greater the mean mega-jam width; when the bottlenecks strength  increases further, 
 only the mega-jam survives and synchronized flow remains only within 
 its downstream front separating free flow and congested traffic. 
 Theoretical results presented explain empirical complexity of traffic congestion
  caused by bad weather conditions or accidents.  
  \end{abstract}

\section{Introduction}
 
The physics of freeway traffic congestion is one of the most quickly developed
fields of  complex spatiotemporal systems.
In empirical observations, traffic breakdown (onset of congestion) in  free flow  
 occurs mostly at  bottlenecks associated with, e.g., 
on- and off-ramps. 
 In congested traffic, moving  jams are    observed~\cite{May,Leu,Gartner,Daganzo,Gazis,Ch,Helbing,Nagatani,Nagel,Mahnke,Maerivoet,KernerBook}.
 A moving  jam is 
 a localized structure of great vehicle density, 
spatially limited by two jam fronts; the jam propagates upstream; within the jam vehicle  speed is very low.
 
 Moving jams, which exhibit a characteristic jam feature [J] to propagate through bottlenecks
  while  maintaining the mean velocity of the downstream jam front, are called
{\it wide} moving jams~\cite{KernerBook}.  A wide moving 
jam  consists of alternations of regions in which traffic flow is 
interrupted and  moving blanks~\cite{KKH,KKHR}; a moving blank is a blank between vehicles, which moves upstream due to vehicle motion within the jam.
The flow interruption can be used as a microscopic criterion for a wide moving jam
associated with the jam definition [J]~\cite{KKH,KKHR}.

In observations~\cite{KernerBook},
traffic breakdown  is associated with 
a local first-order phase transition from free flow to synchronized flow (F$\rightarrow$S transition)
at the bottleneck; 
synchronized flow [S] is defined  
  as congested traffic that  does not exhibit the   feature [J];
   in particular, the downstream front of synchronized flow
is often  \emph{fixed} at the bottleneck. 
  Wide moving jams can emerge spontaneously in synchronized flow 
(S$\rightarrow$J transition) only, i.e.,
due to a sequence of   F$\rightarrow$S$\rightarrow$J transitions.

Moving jam emergence   in synchronized flow leading to 
 S$\rightarrow$J transitions is called the pinch effect occurring within   
an associated pinch region of synchronized flow (see Sect. 12.2 of~\cite{KernerBook}):
 (i) The density increases and  average speed decreases;
narrow moving jams, which do not exhibit the   feature [J],
 appear in the pinch region. (ii)
 These jams propagate upstream growing in their amplitude; as a result,    S$\rightarrow$J transitions occur,
 i.e., wide moving jam emerge.  
   The upstream boundary of the pinch region is a road location at which
   a narrow moving jam has just transformed into a wide moving one.
   (iii) These locations 
    can vary  for different wide moving jams, i.e., the pinch region width 
    depends on time.  A congested traffic pattern
  that exhibits this {\it regular}  structure
   is called a general pattern (GP).

 Earlier traffic flow theories and models reviewed in~\cite{May,Leu,Gartner,Daganzo,Gazis,Ch,Helbing,Nagatani,Nagel,Mahnke} cannot explain
 F$\rightarrow$S$\rightarrow$J transitions and the pinch effect  
 (see a criticism of the  theories  in Ref.~\cite{KernerBook,KKl2006C}).
 For this reason, the author introduced a three-phase traffic theory
 (references in~\cite{KernerBook}) in which there are
 (i) the free flow, (ii) synchronized flow, and (iii) wide moving jam   phases.
The    synchronized flow and wide moving jam phases associated with congested traffic are defined via 
the empirical definitions [S] and [J], respectively. 
The first  three-phase traffic flow models showing the pinch effect
 are stochastic microscopic models~\cite{KKl,KKW}. Later, other three-phase traffic flow models were developed~\cite{Davis,Lee_Sch2004A,Jiang2004A,KKl2006C,Gao2007,Laval2006A}.

 In general, the average flow rate $q^{\rm (cong)}$~\footnote{The  time interval  of flow rate averaging
 is suggested to be 
 considerably longer than  time distances between any moving jams within a congested pattern.} within a congested traffic pattern upstream of a bottleneck
 is the smaller, the greater the bottleneck influence of traffic (called  the bottleneck strength).
 Empirical data show that the flow rate $q^{\rm (cong)}$ 
 within GPs occurring at usual bottlenecks like on- and off-ramp bottlenecks, which is equal to the average flow rate
 in the pinch region $q^{\rm (pinch)}$, is approximately within a  range
   \begin{equation}
 q^{\rm (cong)}= q^{\rm (pinch)}= 1100 - 1700  \ {\rm vehicles/h/lane}.
  \label{pinch}
  \end{equation}
 
Features of GPs and other congested patterns occurring at usual bottlenecks determined by road infrastructure
  (on- and off-ramps, road gradients, etc.), for which condition
 (\ref{pinch}) is valid, have already been studied in detail~\cite{KernerBook}.
 In contrast with the usual bottlenecks,
  due to bad weather conditions or accidents heavy bottlenecks can occur,
  which exhibit a much greater influence on traffic (greater bottleneck strength) that
  limits $q^{\rm (cong)}$ to very small values, sometimes as low as  zero.
  Features of traffic congestion at the heavy bottlenecks are unknown.
  
As follows from recent empirical studies of sequences of wide moving jams~\cite{KKH,KKHR},  the flow rate $q^{\rm (blanks)}$ of
low speed states  associated with moving blanks within the jams
    is approximately within a range
    \begin{equation}
 q^{\rm (blanks)}=  300 - 600  \ {\rm vehicles/h/lane}.
  \label{blanks}
  \end{equation} 
 The flow rate {\it between} the jams associated with non-interrupted flows in the jams' outflows  $q^{\rm (J)}_{\rm out}$
 is  greater than $q^{\rm (pinch)}$ (\ref{pinch}), i.e., $q^{\rm (J)}_{\rm out}$
 is  considerably greater than $q^{\rm (blanks)}$ (\ref{blanks}).
 Now we assume  that due to bad weather conditions or an accident
a     heavy bottleneck occurs with a great strength for which 
\begin{equation}
q^{\rm (cong)}\rightarrow  q^{\rm (blanks)}.
  \label{pinch_wide}
  \end{equation} 
In this case,  $q^{\rm (J)}_{\rm out}$ must reduce to  $q^{\rm (blanks)}$ (\ref{blanks}), i.e.,
the difference between flows within and outside wide moving jams  
    disappears. As a result, all wide moving jams should merge into one mega-wide moving jam
   (mega jam for short).
 Thus already from this qualitative consideration,
  we can  expect   interesting physical phenomena associated with
   complexity of traffic congestion at heavy bottlenecks.

 In this letter, we reveal these features and   compare them with
 measured congested patterns   at 
  heavy   bottlenecks caused by bad weather conditions or accidents.
 
 \section{A Theory of Traffic Congestion at Heavy Bottlenecks}

   For a theoretical analysis of traffic congestion at heavy bottlenecks, 
  we use a stochastic three-phase traffic flow model of a two-lane road~\cite{KKl,KKl2003A}
 (see Appendix)  with the following heavy bottleneck model. 
  
  We suggest that there is a section of the road   
  within which due to an accident or bad weather conditions  
 drivers should increase a safety time gap  
  $\tau^{\rm (safe)}$ to the preceding vehicle. 
  We simulate this effect by an increase in $\tau^{\rm (safe)}$   
   to some 
  $\tau^{\rm (safe)}=T_{\rm B}>$ 1 sec in comparison with  $\tau^{\rm (safe)}=$ 1 sec used in the model 
  for vehicles moving outside this section.  
  In according with  Eq. (\ref{Safety}) of the model (Appendix), $\tau^{\rm (safe)}$     determines
  a safe speed, which should not be exceeded by a driver; otherwise, the driver decelerates. As a result, within this section
  drivers move with   time headways that are approximately equal to $\tau^{\rm (safe)}=T_{\rm B}$.
   Therefore the section with longer $\tau^{\rm (safe)}=T_{\rm B}$ acts  
   as a bottleneck on the road. In the bottleneck model suggested here, each chosen value $T_{\rm B}$ defines a specific bottleneck:
   the strength of this bottleneck  is the greater, the longer $T_{\rm B}$; in turn,  
  the longer    $T_{\rm B}$, the smaller  $q^{\rm (cong)}$, i.e.,
  the greater the flow rate limitation within congestion caused by the bottleneck. This model feature allows us to simulate
  a heavy bottleneck caused by bad weather conditions or accidents  leading to  a long enough  $T_{\rm B}$ 
  within the bottleneck~\footnote{Under bad weather conditions, a road  section
  with a long value $T_{\rm B}$ can be caused, e.g., by 
  a poor view due to  fog on the   section or a much longer deceleration way needed  by snow and ice  
  on the section. If an accident occurs on a road, a road  section
  with a long value $T_{\rm B}$ can be caused, e.g., by 
  much  narrower lane widths allowed for driving on the road section; the same
  effect can occur under heavy roadworks.}.

\begin{figure}
\begin{center}
\includegraphics[width=13 cm]{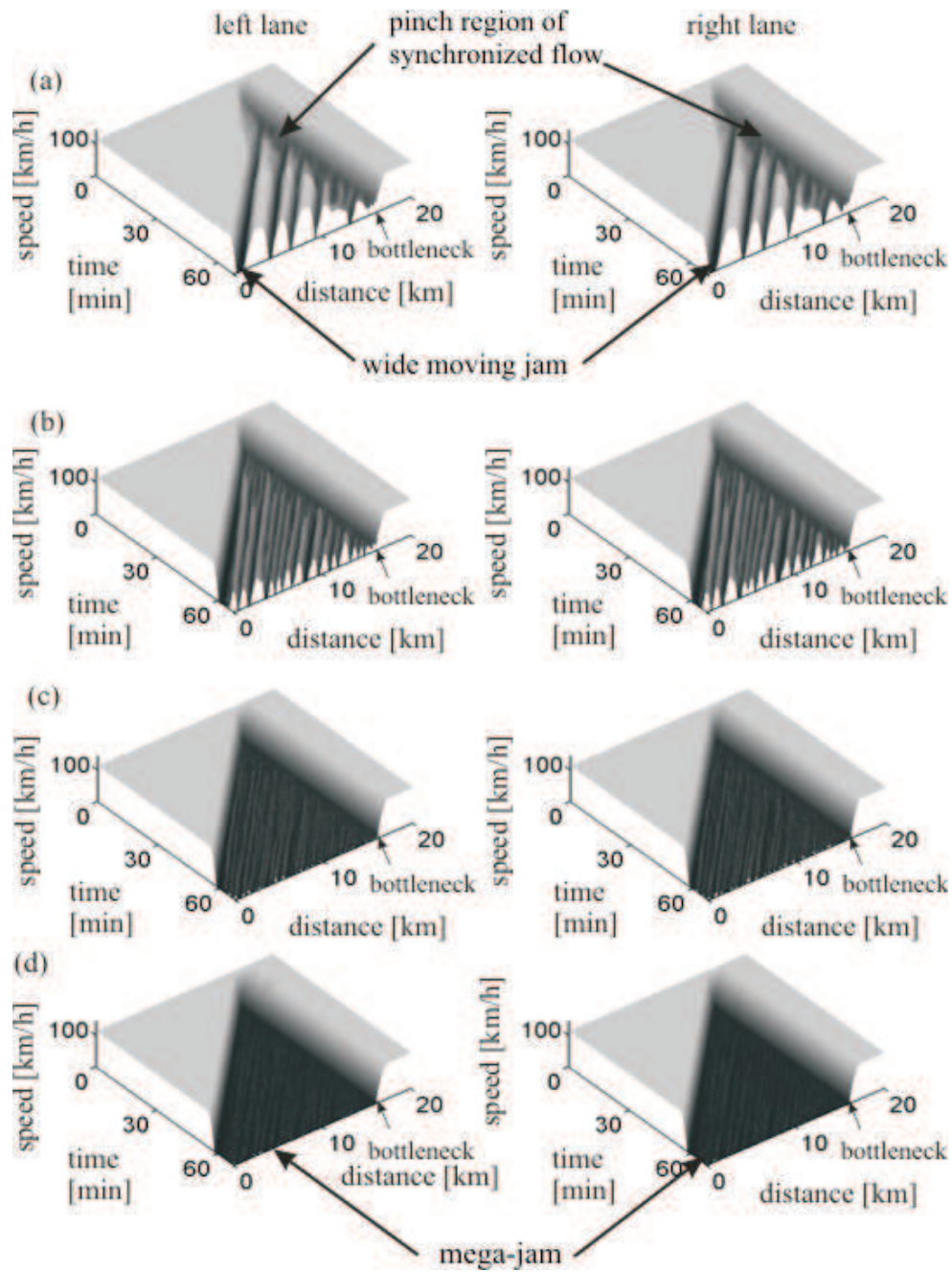}
\caption{Simulated speed in space and time in the left (left) and right (right) road lanes at different $T_{\rm B}$:
$T_{\rm B}=$ 1.8 (a), 2.4  (b), 12 (c), 30 sec (d). $q_{\rm in}=$ 1946 vehicles/h/lane.
The upstream boundary of bottleneck region of the length 300 m is at $x=$ 16 km; the maximum speed wíthin
this region is 60 km/h. Resulting values of
 $q^{\rm (cong)}=$ 1546 (a), 1114 (b),
  440 (c), 217 vehicles/h/lane (d).}
 \label{Mesh_patterns}  
\end{center}
\end{figure}

\begin{figure}
\begin{center}
\includegraphics[width=7 cm]{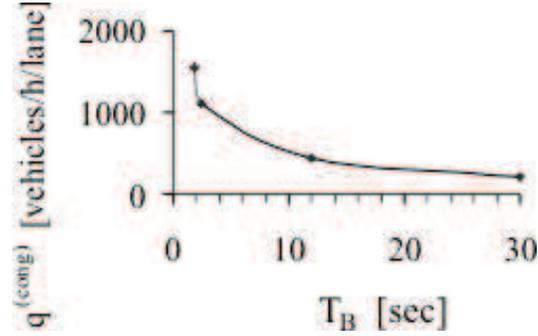}
\caption{Simulated average flow rate $q^{\rm (cong)}$ within   congested patterns shown in Fig.~\ref{Mesh_patterns}
as a function of $T_{\rm B}$ .
The averaging time interval for $q^{\rm (cong)}$ is 60 min.}
 \label{flow}  
\end{center}
\end{figure}

\begin{figure}
\begin{center}
\includegraphics[width=13 cm]{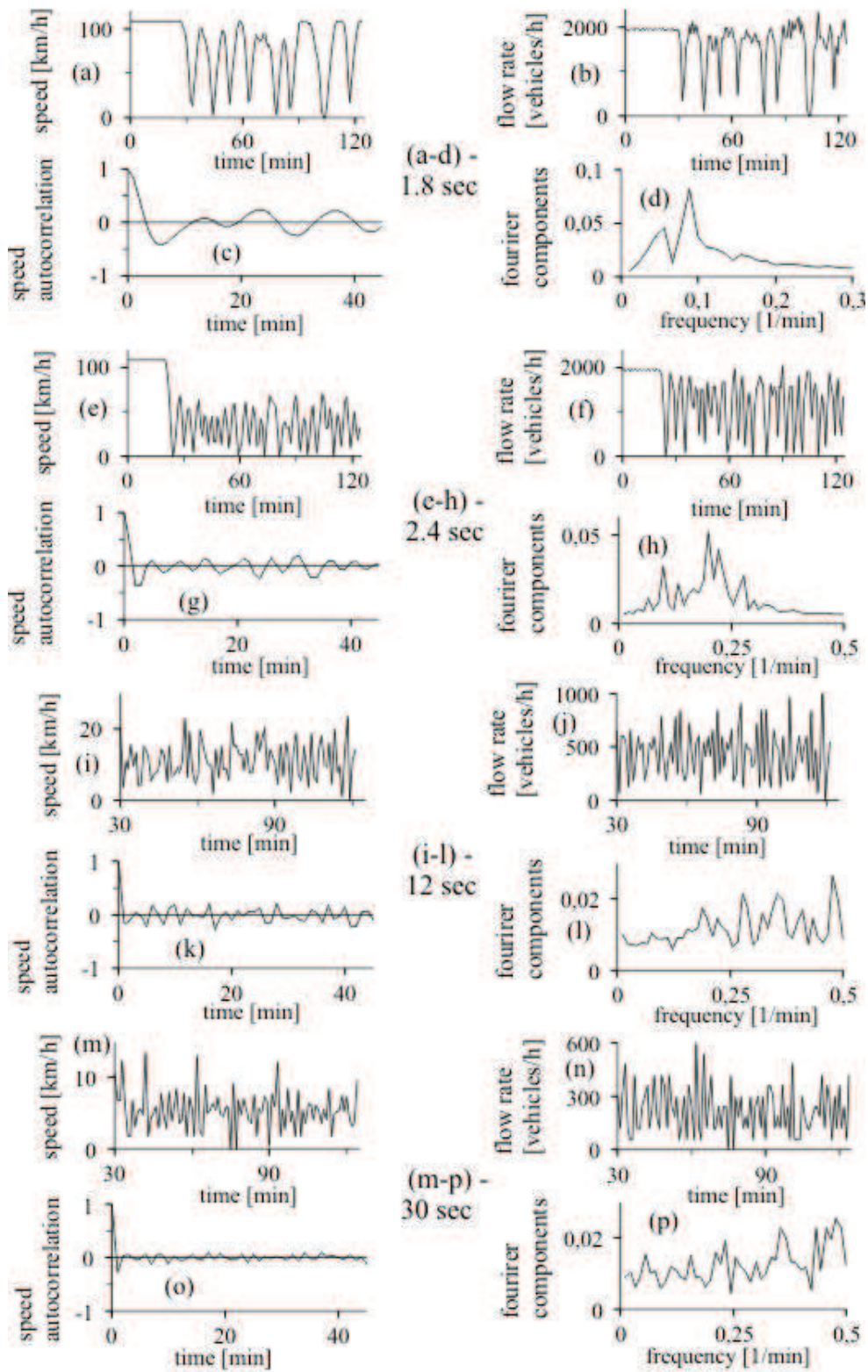}
\caption{Characteristics of congested patterns shown in Fig.~\ref{Mesh_patterns}
 related to location 10 km. Time-functions of speed (a, e, i, m),  speed correlations (c, g, k, o),
  associated Fourier spectra (d, h, l, p), and flow rate (b, f, j, n) for different $T_{\rm B}=$ 
  1.8 (a-d), 2.4  (e-h),  12 (i-l), 30 sec (m-p). 1-min average data.}
 \label{Correlations}  
\end{center}
\end{figure}

\begin{figure}
\begin{center}
\includegraphics[width=13 cm]{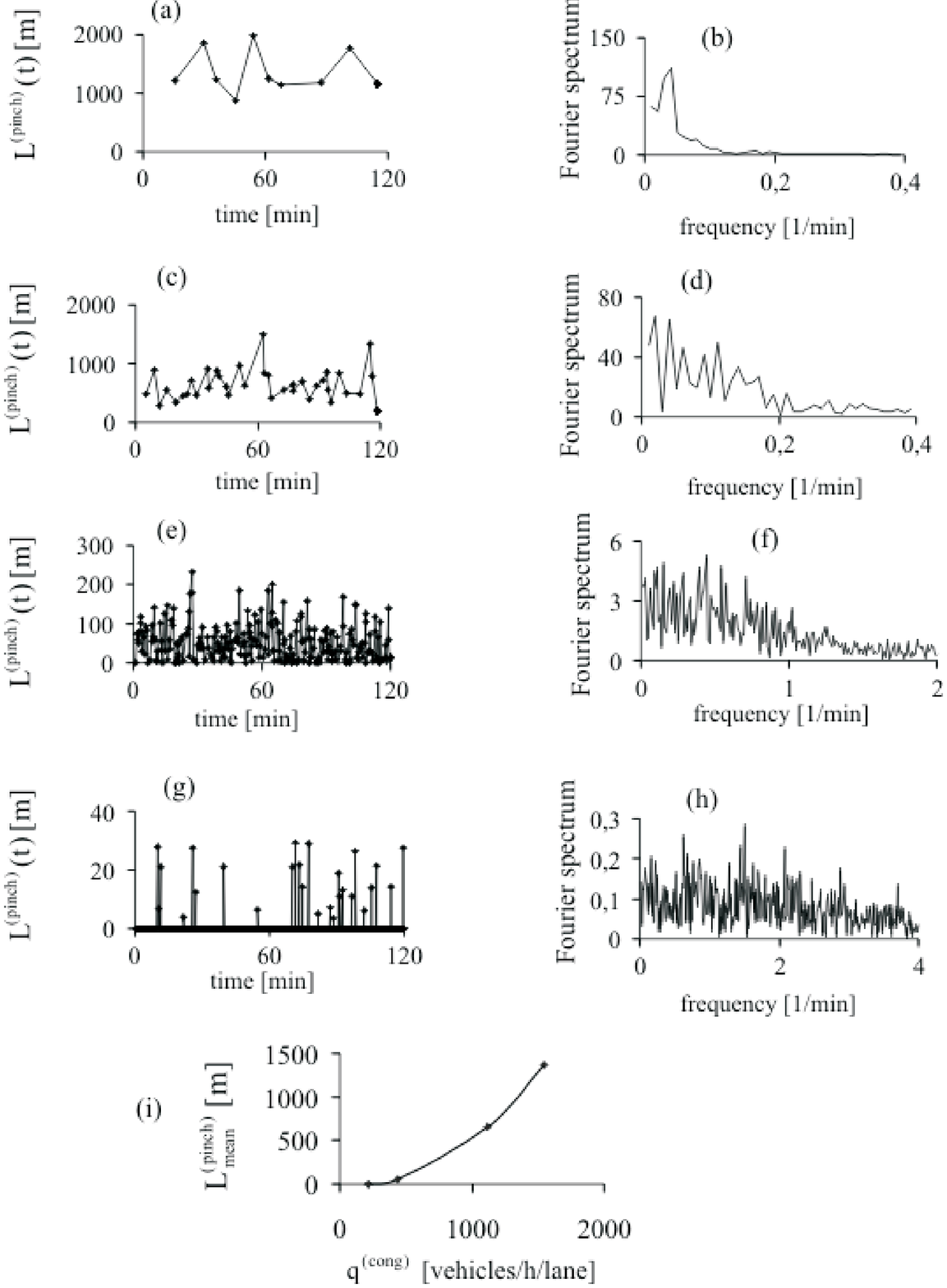}
\caption{Simulations of pinch effect:  $L^{\rm (pinch)}(t)$ (a, c, e, g), associated    Fourier  spectra   
(b, d, f, h) for congested patterns shown in Fig.~\ref{Mesh_patterns},
$L^{\rm (pinch)}_{\rm mean}(q^{\rm (cong)})$ (i). In (a--h)
$T_{\rm B}=$ 1.8 (a, b), 2.4  (c, d),  12 (e, f), 30 sec (g, h). }
 \label{Pinch}  
\end{center}
\end{figure}

  When $T_{\rm B}$ is chosen to be not very long,
  then at a   great enough flow rate $q_{\rm in}$ in free flow upstream of the bottleneck,  firstly an F$\rightarrow$S transition
  occurs spontaneously at the bottleneck. Then the pinch region with a relatively great 
  flow rate $q^{\rm (cong)}=q^{\rm (pinch)}$  is formed.  At the pinch region upstream boundary 
    wide moving jams emerge. Thus we found  known phenomena 
  of regular
  GP formation 
  (Figs.~\ref{Mesh_patterns} (a),~\ref{flow}  and~\ref{Correlations} (a, b))~\cite{KernerBook}. 
  The pinch region width   $L^{\rm (pinch)}(t)$ changes over time between about 1 and 2 km (Fig.~\ref{Pinch} (a)). 
  $L^{\rm (pinch)}$  is defined  as
  the distance between the upstream boundary of the bottleneck  ($x=$ 16 km) and the road location upstream at which a wide moving jam    
  has just been identified through the use of the jam microscopic criterion.
  There is also a nearly constant frequency of $L^{\rm (pinch)}(t)$ oscillations associated with the maximum
  in the Fourier spectrum  (Fig.~\ref{Pinch} (b)). Speed autocorrelation functions and associated Fourier spectra of  speed time-dependencies
  at shorter $T_{\rm B}$  show regular character of wide moving jam propagation
  (Fig.~\ref{Correlations} (c, d)).
  
    When $T_{\rm B}$ increases  and therefore
    $q^{\rm (cong)}=q^{\rm (pinch)}$  decreases    (Fig.~\ref{flow}), and $T_{\rm B}$ remains a relatively small value
    ($1.8< T_{\rm B}< 3$ sec), then
     as for other   bottleneck types, 
    we found the following known GP features~\cite{KernerBook}: 
   the smaller  the flow rate $q^{\rm (pinch)}$, the greater the frequency
  of narrow moving jam emergence within the pinch region, the lower the maximum speed
  between wide moving jams upstream of the pinch region,   the smaller the mean pinch region width
  $L^{\rm (pinch)}_{\rm mean}$
  (Figs.~\ref{Mesh_patterns} (b) and~\ref{Pinch} (c, d, i)). 
  This can also be seen from a comparison of time-dependences of average speed  
  within the region of wide moving jams
  for different $T_{\rm B}$ (Fig.~\ref{Correlations} (a, e)).

  Qualitatively other  phenomena are found when   $T_{\rm B}$   further increases
($T_{\rm B}> 3$ sec)  and the average flow rate    $q^{\rm (cong)}$  decreases     
   considerably 
   (Figs.~\ref{Mesh_patterns} (c, d) and~\ref{flow}).
   Firstly,
   $L^{\rm (pinch)}(t)$ becomes  a non-regular time-function (Fig.~\ref{Pinch} (e, g)) whose Fourier spectrum
  is broader, the longer $T_{\rm B}$ (Fig.~\ref{Pinch} (f, h)); there are random time intervals when the pinch region   disappears,
  i.e., $L^{\rm (pinch)}= 0$ (Fig.~\ref{Pinch} (e)).
 This means that there are time instants at which there is no pinch region and wide moving jams emerge directly
  at the upstream boundary of the bottleneck, whereas for other time intervals
  the pinch region appears again (Fig.~\ref{Pinch} (e)). 
 Consequently,  $L^{\rm (pinch)}_{\rm mean}$ decreases continuously, when $T_{\rm B}$ increases
  (Fig.~\ref{Pinch} (i)).
   $L^{\rm (pinch)}_{\rm mean}$ reaches almost zero for $T_{\rm B}=$ 30 sec:
   for such a heavy   bottleneck, the pinch region 
    does not exist.

   When $T_{\rm B}$ increases and 
   the pinch region disappears during some random time intervals,
  the GP   behaviour   qualitatively changes:
  the time-functions of average speed   become non-regular ones (Fig.~\ref{Correlations} (i, m)).
  We found that under condition (\ref{pinch_wide}), the pinch region 
    disappears and   wide moving jams merge into a  mega-jam. 
In particular,
   for
  $T_{\rm B}=$ 30 sec
  no separated wide moving jams can be distinguished: instead of the sequence of wide moving jams, a  mega-jam 
  occurs {\it only}.   These effects can also be seen from speed autocorrelation functions and associated Fourier spectra of the average speed time-dependences
  (Fig.~\ref{Correlations} (k, l, o, p)). 
  The mega-jam  is found at those long values of $T_{\rm B}$ at which
  the pinch region disappears. 
   Simulations show that the  mega-jam    
    consists of alternations of flow interruption and low speed states associated with moving blanks. Thus there are no
    continuous flows within the congested pattern any more with the   one exception of the  downstream front of synchronized flow
    that separates free flow downstream and the mega-jam upstream of the front.
    Synchronized flow remains {\it only} within this front.
 Thus in accordance with   
     all known empirical data and theoretical results presented,   the   condition 
  \begin{equation}
 q^{\rm (cong)} >q^{\rm (blanks)}
  \label{nec}
  \end{equation}
is a {\it necessary} condition for GP existence at a bottleneck.

  For explanations of these results, we recall
   that an S$\rightarrow$J transition is a first-order phase transition~\cite{KernerBook}, i.e., it
  is characterized by a {\it random time delay} $T_{\rm SJ}$.  $T_{\rm SJ}$ 
  includes a random time
  delay of spontaneous nucleation of a narrow moving jam (nucleus for the S$\rightarrow$J transition)
      and its random growth within the pinch region 
   until the jam transforms into a wide moving jam
  at the upstream boundary of the pinch region. Thus the smaller $T_{\rm SJ}$, the shorter  $L^{\rm (pinch)}$. The random character
  of   $T_{\rm SJ}$  explains
 a time-dependence
  of   $L^{\rm (pinch)}(t)$ (Fig.~\ref{Pinch} (c, e)). We found that the greater $T_{\rm B}$, the greater the density within the pinch region
  and the smaller the mean random time delay $T^{\rm (mean)}_{\rm SJ}$. 
  Thus the  greater $T_{\rm B}$, the greater the probability for the occurrence of a  negligibly small $T_{\rm SJ}$
  and   wide moving jam emergence  directly upstream of the   synchronized flow front; in the latter case 
   $L^{\rm (pinch)}=$ 0. 
This   explains also
   why $L^{\rm (pinch)}_{\rm mean}$ decreases up to zero, when an increase in $T_{\rm B}$ causes 
  a decrease in $T^{\rm (mean)}_{\rm SJ}$ up to zero, i.e., when all wide moving jams
   emerge  almost directly upstream of the bottleneck and the pinch region disappears.

 When  
      $q^{\rm (cong)}$ becomes zero, because behind a road location  
    the   road is    closed, 
      the mega-jam transforms into a 
     queue of {\it motionless} vehicles,
      which therefore is not associated with vehicular traffic.  Nevertheless, there is a link between the queue and the mega-jam.
     If at a time instant
     we allow  several vehicles   to escape from
    this queue, then simulations show that  motion of these vehicles results in   wide moving jam occurrence: the downstream front of the jam
     separates  moving vehicles escaping from the queue
    and vehicles standing within the queue upstream of the front. When the number of vehicles escaping from the
    initial queue decreases, the downstream jam front transforms into a moving blank(s)  
    subsequently covered by  vehicles standing   in the queue.
      When a vehicle per a long enough time interval is allowed to escape  from the mega-jam,
     as simulations show,
     a sequence of  such moving blanks within this jam occur; these moving blanks  exhibit, however,    
     dynamics   specifically
     associated with     the mega-jam. 
     
     Thus based on a study of three-phase traffic flow model
     we found that the complexity of traffic congestion caused by bad weather conditions or accidents, when 
a very heavy bottleneck appears on a road, is associated with the phenomenon of random disappearance and appearance of the pinch region over time that is accompanied by the random merger of some wide moving jams into a mega-jam. When the bottleneck strength increases further strongly, the pinch region disappears and only the mega-jam survives and synchronized flow remains only within its downstream front separating free flow and congested traffic. These phenomena explained by random nucleation of moving jams in congested traffic reveal the evolution of the traffic phases when heavy bottlenecks occur in highway traffic.
 
 \section{Discussion. Comparison with Empirical Results}
  
\begin{figure}
\begin{center}
\includegraphics[width=8 cm]{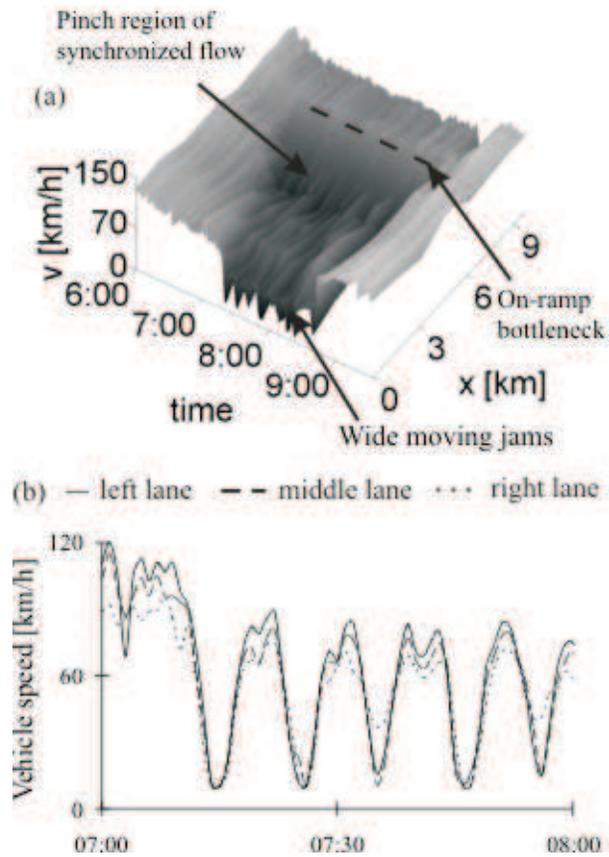}
\caption{Usual   empirical traffic congestion -- a general pattern (GP) at on-ramp bottleneck:
 (a)   Speed in space and time for the GP. (b) Speed in different lanes within a sequence of wide moving jams associated with (a) at location 1.5 km. 
1-min average data measured on the freeway A5-South in Germany dated 
April 15, 1996. Arangement of detectors is shown in Fig. 2.1 of the book~\protect\cite{KernerBook}.}
 \label{emp_GP}  
\end{center}
\end{figure}

\begin{figure}
\begin{center}
\includegraphics[width=17 cm]{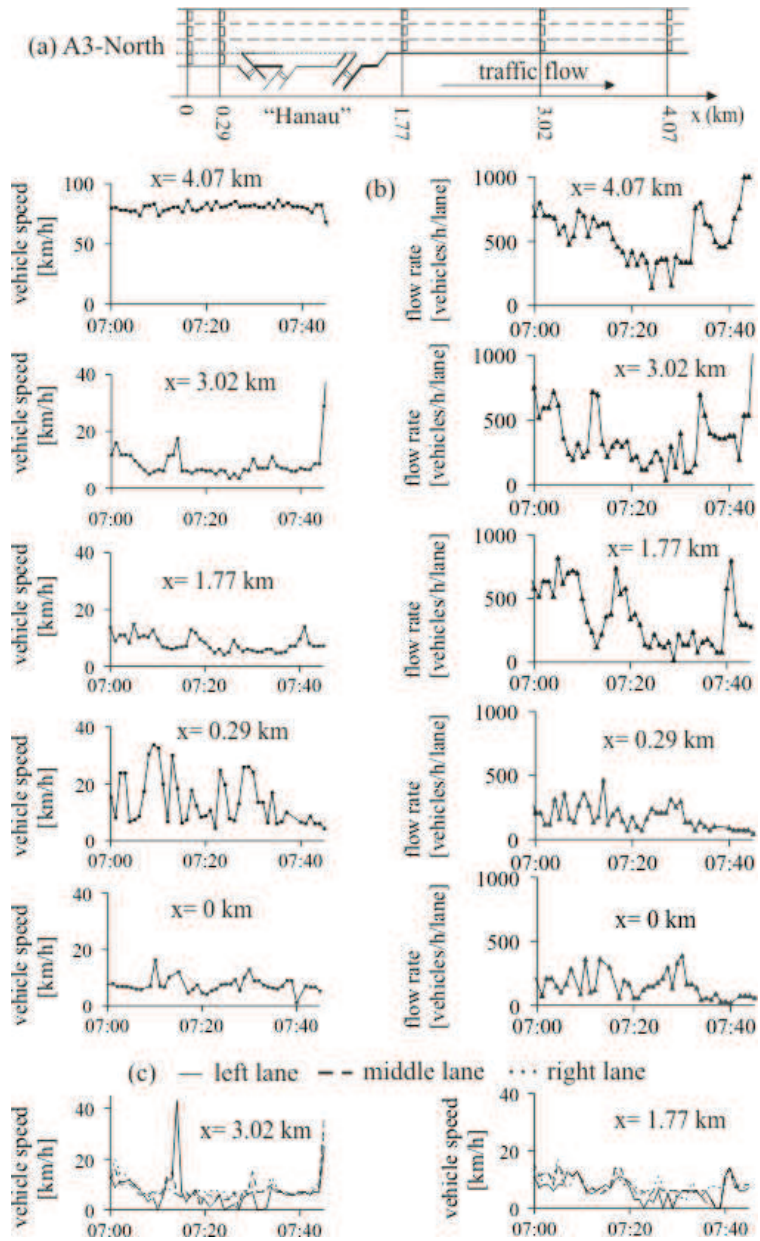}
\caption{Empirical structure of congestion caused by snow and ice:
(a) Schema of road detector arrangement on a section of the  freeway A3-North in Germany near the intersection
$\lq\lq$Hanau". (b) Average speed and flow rate across the freeway
  at different locations. (c) Speed in different lanes at two locations.
1-min average data.}
 \label{emp_06022006}  
\end{center}
\end{figure}

 To compare the above theory of traffic congestion   with   empirical 
 congested patterns, one should have measured data for traffic congestion at a bottleneck, whose strength
 should be manually continuously changeable from the one associated with usual
 bottlenecks like on- and off-ramps to   great bottleneck strengths associated with heavy bottlenecks caused by
 bad
 weather conditions or accidents. Unfortunately, such measured data is not available. However, we can
 compare the theory with measured data related to two limiting cases:
 (i) Traffic congestion at an usual on-ramp bottleneck  (Fig.~\ref{emp_GP}). (ii) Traffic congestion at a heavy bottleneck caused by
 bad weather conditions -- snow and ice on a road (Fig.~\ref{emp_06022006}).

 The  speed distributions with congested patterns found in simulations (Fig.~\ref{Mesh_patterns} (a, b))
  for the range of $T_{\rm B}=1.6-2.4$   sec associated with the flow rate range   
    \begin{equation}
 q^{\rm (cong)}= q^{\rm (pinch)} =1120 - 1800 \ {\rm vehicles/h/lane}
  \label{pinch_t}
  \end{equation}
  are qualitatively     the same as those 
  in an  empirical GP shown in Fig.~\ref{emp_GP}  and
  in all other known empirical GPs~\cite{KernerBook}. Moreover,
   quantitative   values of   empirical flow rates $q^{\rm (pinch)}$ (\ref{pinch})
   are approximately associated with the theoretical result (\ref{pinch_t}). In particular, within the pinch region 
  of  the GP shown in Fig.~\ref{emp_GP} (a), the flow rate averaged between 7:00 and 8:00 and across the road is 
  $q^{\rm (pinch)}=1200$ vehicles/h/lane.
  
  In contrast, for $T_{\rm B}\geq $ 6 sec 
  associated with  the average flow rate
    \begin{equation}
 q^{\rm (cong)} \leq 625 \ {\rm vehicles/h/lane}
  \label{blanks_t}
  \end{equation}
  the theoretical speed  patterns (Fig.~\ref{Mesh_patterns} (c, d))
  are qualitatively the same as those we found in empirical data measured on many various days (and years)
  on different freeways, when
   very heavy bottlenecks caused by bad weather conditions or accidents occur
   for which empirical flow rates  $q^{\rm (cong)}< 700$ vehicles/h/lane. In this case, 
   rather than regular structure of traffic congestion of   GPs (Figs.~\ref{Mesh_patterns} (a, b)
   and~\ref{emp_GP}), both
       empirical and theoretical traffic congested patterns (Figs.~\ref{Mesh_patterns} (c, d)
   and~\ref{emp_06022006}) exhibit 
    non-regular   spatiotemporal structure of congestion in which no sequences of wide moving jams
    can be distinguished.
  In addition,   the mentioned  empirical flow rates  $q^{\rm (cong)}< 700$ vehicles/h/lane
 found within the empirical non-regular congested patterns
   are associated with the theoretical result (\ref{blanks_t}). In turn, these theoretical and empirical results
   for $q^{\rm (cong)}$ correspond
    to   empirical flow rates found within wide moving jams associated with moving blanks (\ref{blanks}).
     
 An example of such an empirical congested traffic pattern
  is shown in Fig.~\ref{emp_06022006}.    
  Indeed, rather than   the regular  structure of congestion within the GP (Fig.~\ref{emp_GP}), in    
     measured data associated with  bad weather conditions  
a  non-regular   spatiotemporal structure of congestion  is  observed (Fig.~\ref{emp_06022006}). A heavy   bottleneck appears
on February 02, 2006 between  locations
4.07 and 3.02 km due to snow and ice. Upstream of the bottleneck, very low speed and flow rate patterns
 ($x\leq$ 3.02 km in Fig.~\ref{emp_06022006} (b))
are observed. Downstream of the bottleneck ($x=$ 4.07 km) vehicles have escaped from the congestion (speed is high), however, the flow rate is very small
because the bottleneck reduced
the average flow rate within the congestion  strongly. For example, 
at $x=$ 3.02 km, the flow rate  $q^{\rm (cong)}$ averaged between 7:00 and 7:40 and across
the road  
is only   513 vehicles/h/lane that corresponds to the theoretical result (\ref{blanks_t}). In contrast with the GP in
Fig.~\ref{emp_GP} (b) (as with other empirical GPs~\cite{KernerBook}),
within traffic congestion shown in Fig.~\ref{emp_06022006} (b) non-regular low speed patterns 
are observed in which a sequence of wide moving jams cannot be distinguished  ($x\leq$ 3.02 km). 
This conclusion is regardless of the flow rates to on- and off-ramps, which  
in the data set lead to a greater average flow rate downstream of the freeway intersection    
(Fig.~\ref{emp_06022006} (a, b)).

 If speeds in different lanes are compared (Fig.~\ref{emp_06022006} (c)), we   find even   more non-regular
speed time-dependencies: whereas  no  vehicles    pass a detector in one of the lanes, i.e., the speed is zero,
 during the same time interval  the average speed in  other lanes can be higher than zero.
 This explains why in time-dependences of speeds averaged across the road (Fig.~\ref{emp_06022006} (b)),
 this very low speed is seldom equal to exactly  zero.

I   thank Sergey Klenov for help in simulations, 
Ines Maiwald-Hiller for help in data analysis, Andreas Hiller, 
Gerhard N\"ocker, Hubert Rehborn and Olivia Brickley for suggestions.

  \appendix

\section{Stochastic Microscopic Three-Phase Traffic Flow Model}
\label{A}

The stochastic microscopic three-phase traffic flow model for two-lane road
used in all simulations presented above
reads as follows~\cite{KKl,KKl2003A}:
\begin{equation}
v_{n+1}=\max(0, \min({v_{\rm free}, \tilde v_{n+1}+\xi_{n}, v_{n}+a \tau, v_{{\rm s},n} })),
\label{final}
\end{equation}
\begin{equation}
\label{next_x}
x_{n+1}= x_{n}+v_{n+1}\tau,
\end{equation}
\begin{eqnarray}
\label{next}
\tilde v_{n+1}=\max(0, \min({v_{\rm free}, v_{{\rm c},n}, v_{{\rm s},n} })), \\
 \\
v_{c,n}=\left\{
\begin{array}{ll}
v_{n}+\Delta_{n} &  \textrm{at $g_{n}
\leq G_{n}$} \\
v_{n}+a_{n}\tau &  \textrm{at $g_{n}> G_{n}$},
\end{array} \right.
\label{next1}
\end{eqnarray}
where
\begin{equation}
\Delta_{n}=\max(-b_{n}\tau, \min(a_{n}\tau, \ v_{\ell,n}-v_{n})),
\label{next2}
\end{equation}  
where
index $n$ corresponds to the discrete time $t=n\tau$,
$n=0,1,2,...;$ $\tau$ is time step; $v_{\rm n}$
is the vehicle speed at time step $n$; $x_{n}$ is the vehicle co-ordinate;
$\tilde v_{\rm n}$ is the speed calculated without a noise component $\xi_{n}$;
$v_{\rm free}$ is the maximum speed 
in free flow that is  constant;  $a$ is the maximum acceleration;
 $g_{n}=x_{\ell, n}-x_{n}-d$ is the space gap; the lower index $\ell$
marks functions and values related to the preceding vehicle;
$v_{\rm s,n}$ is a safe speed; 
all vehicles have the same length $d$; $a_{n}\geq 0$,  $b_{n}\geq 0$ (see below); 
$G_{n}$ is a synchronization gap:
\begin{equation}
G_{n}=G(v_{n}, v_{\ell,n}),
\label{D}
\end{equation}
where the function $G(u, w)$ is chosen as
\begin{equation}
G(u, w)=\max(0, k\tau u+ \phi_{0} a^{-1}u(u-w)),
\label{D_function}
\end{equation}
$k>1$ and $\phi_{0}$ are constants.

The safe speed $v_{{\rm s},n}$ in (\ref{final}) is chosen in the form 
\begin{equation}
v_{{\rm s},n}=
\min{(v^{\rm (safe)}_{ n},  g_{ n}/ \tau+ v^{\rm (a)}_{ \ell})},
\label{safe}
\end{equation}
where $v^{\rm (safe)}_{ n}=v^{\rm (safe)}(g_{ n},v_{ \ell,n})$~\cite{Kra} 
 is a solution of the
 Gipps-equation~\cite{Gipps} 
\begin{equation}
v^{\rm (safe)} \tau^{\rm (safe)} + X_{\rm d}(v^{\rm (safe)}) = g_{n}+X_{\rm d}(v_{\rm \ell, n}),
\label{Safety}
\end{equation}
where $X_{\rm d}(u)$ is the distance travelled by the vehicle with an initial speed $u$ at
a time-independent deceleration $b$
until it
comes to a stop; 
$v^{\rm (a)}_{ \ell}$  is an $\lq\lq$anticipation" speed of the preceding vehicle at the 
next time step (formula (16.48) of~\cite{KernerBook}).

 The noise component   $\xi_{n}$ in (\ref{final})
 that simulates random deceleration and acceleration is applied depending on whether the vehicle
 decelerates or accelerates, or else maintains its speed:
 \begin{equation}
\xi_{n}=\left\{
\begin{array}{ll}
-\xi_{\rm b} &  \textrm{if $S_{n+1}=-1$} \\
\xi_{\rm a} &  \textrm{if  $S_{n+1}=1$} \\
0 &  \textrm{if  $S_{n+1}=0$},
\end{array} \right.
\label{noise}
\end{equation}
where  
$S$ in (\ref{noise})  denotes the state of motion ($S_{n+1}=-1$ represents
   deceleration, $S_{n+1}=1$    acceleration, and $S_{n+1}=0$  
motion at nearly constant speed)
\begin{equation}
S_{n+1}=\left\{
\begin{array}{ll}
-1 &  \textrm{if $\tilde v_{n+1}< v_{n}-\delta$} \\
1 &  \textrm{if $\tilde v_{n+1}> v_{n}+\delta$} \\
0 &  \textrm{otherwise},
\end{array} \right.
\label{noise_}
\end{equation}  
$\delta$ is a constant 
($\delta \ll a\tau$),
\begin{equation} 
 \xi_{\rm a}=a\tau \Theta (p_{\rm a}-r),
 \label{xi_acc} 
 \end{equation}
  \begin{equation} 
 \xi_{\rm b}=a\tau \Theta (p_{\rm b}-r),
 \label{xi_dec} 
 \end{equation} 
  where  $p_{\rm a}$ and $p_{\rm b}$ are probabilities of random acceleration and deceleration, respectively;
  $r={\rm rand (0,1)}$,
 $\Theta (z) =0$ at $z<0$ and $\Theta (z) =1$ at $z\geq 0$. 
 
 To simulate driver time delays in acceleration or deceleration,   $ a_{n}$ and  $ b_{n}$
 are taken as stochastic functions
 \begin{equation} 
 a_{n}=a \Theta (P_{\rm 0}-r_{1})
 \label{r_acc} 
 \end{equation}  
 \begin{equation} 
 b_{n}=a \Theta (P_{\rm 1}-r_{1}),
 \label{r_dec} 
 \end{equation} 
\begin{equation}
P_{0}=\left\{
\begin{array}{ll}
p_{0}(v_{n}) & \textrm{if $S_{n} \neq 1$} \\
1 &  \textrm{if $S_{n}= 1$},
\end{array} \right.
\label{prob0}
\end{equation}
\begin{equation}
P_{1}=\left\{
\begin{array}{ll}
p_{1} & \textrm{if $S_{n}\neq -1$} \\
p_{2}(v_{n}) &  \textrm{if $S_{n}= -1$},
\end{array} \right.
\label{prob1}
\end{equation}
where speed functions for probabilities $p_{0}(v_{n})$ and $p_{2}(v_{n})$ are considered
in~\cite{KernerBook}; $p_{1}$ is constant; $r_{1}={\rm rand (0,1)}$. 

The following  incentive
conditions
 for  lane changing from
the right lane to the left lane ($R \rightarrow L$) and a return change from the left lane
to the right lane ($L \rightarrow R$) have been used in the model:
\begin{equation}
R \rightarrow L: \ v^{+}_{n} \geq v_{\ell, n}+\delta_{1} \ \textrm{and} \ v_{n}\geq v_{\ell, n},
\label{RLi}
\end{equation}
\begin{equation}
L \rightarrow R: \ v^{+}_{n} > v_{\ell, n}+\delta_{1} \ \textrm{or} \ v^{+}_{n}>v_{n}+\delta_{1}.
\label{LRi}
\end{equation}
The  security conditions for  lane changing are given by the inequalities:
\begin{equation}
g^{+}_{n} >\min(v_{n}\tau , \ G^{+}_{n}),  
\label{S1}
\end{equation}
\begin{equation}
g^{-}_{n} >\min(v^{-}_{n}\tau, \ G^{-}_{n}), 
\label{S2}
\end{equation}
where 
\begin{equation}
G^{+}_{n}=G( v_{n}, v^{+}_{n}),
\label{Dplus}
\end{equation}
\begin{equation}
 G^{-}_{n}=G(v^{-}_{n}, v_{n}),
\label{Dminus}
\end{equation}
 the function  $G(u, w)$
is given by  (\ref{D_function});
 the speed  $v^{+}_{n}$ or the speed $v_{\ell, n}$ in (\ref{S1}), (\ref{S2}) is 
set to $\infty$
if the 
space gap $g^{+}_{n}$ or the space gap $g_{n}$
exceeds a given look-ahead distance $L_{\rm a}$, respectively;
superscripts    $+$   and  $-$  in variables, parameters, and functions 
denote the preceding vehicle and the trailing vehicle 
in the $\lq\lq$target" (neighboring) lane, respectively. The target lane is the 
lane into which the vehicle wants to change. If the conditions (\ref{RLi})--(\ref{S1}) are satisfied,  
the vehicle changes    lanes with  probability
$p_{\rm c}$. $p_{\rm c}$, $\delta_{1}$ ($\delta_{1} \geq 0$), $L_{\rm a}$ are 
 constants.

Explanations of the  physics of this traffic flow model are given in the book~\cite{KernerBook}.


\begin{thebibliography}{8.}
\addcontentsline{toc}{section}{References}




\bibitem{May}


 May AD    (1990) {\it Traffic Flow Fundamentals}  (Prentice-Hall, Inc., New Jersey)

\bibitem{Leu}  
  Leutzbach W   (1988) {\it Introduction to the Theory of Traffic Flow} (Springer, Berlin).
 	
\bibitem{Gartner}  	
 Gartner NH,  Messer CJ,   Rathi A (eds.)  (1997)
 	{\it Special Report 165: Revised Monograph on Traffic Flow Theory} (Transportation Research Board, Washington, D.C.)
 	
 	\bibitem{Daganzo}  
 	 Daganzo CF (1997)
{\em Fundamentals of Transportation and Traffic Operations}
(Elsevier Science Inc., New York)

\bibitem{Gazis} 
 Gazis DC (2002) {\it Traffic Theory} (Springer, Berlin)

\bibitem{Ch} 
 Chowdhury D,  Santen L,   Schadschneider A  (2000)
{\it Physics Reports} {\bf 329} 199 

\bibitem{Helbing} 
 Helbing D (2001)
{\it Rev. Mod. Phys.} {\bf 73} 1067--1141 

\bibitem{Nagatani} 
 Nagatani T (2002)
{\it Rep. Prog. Phys.}  {\bf 65} 1331--1386 


\bibitem{Nagel} 
  Nagel K,   Wagner P,     Woesler R (2003)
{\it Operation Res.} {\bf 51} 681--716 


\bibitem{Mahnke} 
   Mahnke R,    Kaupu\v{z}s J,  Lubashevsky I (2005) 
{\it Phys. Rep.}
{\bf 408} 1--130  

\bibitem{Maerivoet}

 Maerivoet S,   De Moor B (2005)  {\it Phys. Rep.}  {\bf 419} 1-64

\bibitem{KernerBook}
Kerner BS (2004) {\em The Physics of Traffic} 
(Springer, Berlin, New York) 

 \bibitem{KKH}
Kerner BS,   Klenov SL,    Hiller A (2006)
{\it J. Phys. A: Math. Gen.}  {\bf 39} 2001--2020 

 \bibitem{KKHR}

Kerner BS,   Klenov SL,    Hiller A,   Rehborn H  (2006)
{\it Phys. Rev. E} {\bf 73} 046107  



  

\bibitem{KKl2006C}

Kerner BS and   Klenov SL  (2006)
{\it J. Phys. A: Math. Gen.} {\bf 39} 1775--1809 

  

\bibitem{KKl}

Kerner BS and   Klenov SL  (2002)
 {\it J. Phys. A: Math. Gen.} {\bf 35} L31--L43 
 
 

\bibitem{KKW}

 Kerner BS,   Klenov SL,  Wolf DE  (2002)
{\it J. Phys. A: Math. Gen.} {\bf 35} 9971--10013 
 
 
 \bibitem{Davis}
 
 Davis LC (2004) {\it Phys. Rev. E} {\bf 69} 016108  


 \bibitem{Lee_Sch2004A}

 Lee HK,   Barlovi\'{c} R, Schreckenberg M,    Kim D (2004)
{\it Phys. Rev. Lett.} {\bf 92}, 238702  

\bibitem{Jiang2004A}

Jiang R and  Wu Q-S (2004)
{\it J. Phys. A: Math. Gen.} {\bf 37} 8197--8213 


\bibitem{Gao2007}
 Gao K,   Jiang R,     Hu S-X,   Wang B-H,  Wu Q-S  (2007)
 {\it Phys. Rev. E} {\bf 76}  
 
  \bibitem{Laval2006A} 
  Laval JA (2007)
 In: 
{\it Traffic and Granular Flow' 05}, A. Schadschneider,  T. P\"oschel, R. K\"uhne, 
M. Schreckenberg,  D.E. Wolf  (eds.) 
(Springer, Berlin) p. 521--526

\bibitem{KKl2003A}

Kerner BS and   Klenov SL (2003)
 {\it Phys. Rev. E} {\bf 68} 036130 

 \bibitem{Kra}

  Krau{\ss} S,    Wagner P, Gawron  C  (1997)
{\it Phys. Rev. E}  {\bf 55}  5597--5602

  \bibitem {Gipps}
Gipps PG (1981)
{\it Trans Res B} {\bf 15} 105--111 

  
   
\end{thebibliography}
\end{document}